\documentclass[twocolumn]{aastex631}

\newcommand{\change}[1]{#1}

\usepackage{amsmath}
\usepackage{bm}
\usepackage{CJK}
\usepackage{enumitem}

\begin{document}

\title{The Role of Self-Gravity in Debris Disk Warp Formation: The Case of HD~110058}

\author[0000-0001-8544-0280]{Gang Zhao 
}
\affiliation{Nanjing Institute of Astronomical Optics \& Technology, Chinese Academy of Sciences}
\affiliation{CAS Key Laboratory of Astronomical Optics \& Technology, Nanjing 210042, China}

\author[0000-0002-4859-259X]{Su Wang  
}
\correspondingauthor{Su Wang}
\email{wangsu@pmo.ac.cn}
\affiliation{Purple Mountain Observatory, Chinese Academy of Sciences}
\affiliation{School of Astronomy and Space Science, University of Science and Technology of China, Hefei 230026, China}

\author[0000-0002-7612-6377]{Jiangpei Dou  
}
\affiliation{Nanjing Institute of Astronomical Optics \& Technology, Chinese Academy of Sciences}
\affiliation{CAS Key Laboratory of Astronomical Optics \& Technology, Nanjing 210042, China}



\begin{abstract}
We investigate the crucial role of self-gravity in the formation of warps in debris disks, focusing on the HD~110058 system as an example. Using advanced, GPU-accelerated $N$-body simulations, we model the gravitational dynamics of a massive planetesimal disk perturbed by an inclined, inner planet. Our simulations reveal that self-gravity fundamentally alters the disk's evolution compared to massless models. It enforces a coherent, semi-rigid precession of the disk and enables the rapid formation of a global warp structure within 0.5~Myr. The warp angle undergoes a damped oscillation, eventually settling into a quasi-equilibrium state. By generating synthetic scattered-light images, we demonstrate that our model successfully reproduces the observed S-shaped warp morphology of the debris disk in HD~110058, supporting the existence of an unseen planet. Furthermore, we derive an empirical relationship that connects the equilibrium warp angle to the physical parameters of the disk and the planet. Applying this relation to HD 110058, we constrain its disk mass to be likely less than 1,000~$M_\oplus$, offering a new dynamical perspective on the debris disk mass problem.
\end{abstract}

\keywords{Circumstellar disks (235); Debris disks (363); Exoplanet dynamics (490); N-body simulations (1083)}


\section{Introduction} \label{sec:intro}
Debris disks are the remnants of the planetary formation process, consisting of solid materials from micron-scale dust to kilometer-sized planetesimals \citep{hughes18}. To date, tens of extrasolar debris disks have been resolved with a variety of structures including eccentric rings, gaps, warps, bumps, spiral arms and other asymmetric structures\citep{hughes18}. These structures are likely related to the presence of planets within the systems \citep{nederlander21, kalas05, faramaz19, wyatt06, lagrange25a}. 

For a warped debris disk, although it is widely accepted that it is linked to an inclined planet, the deep dynamical mechanism behind it is still not clear. Early studies \citep{mouillet97, augereau01} on the $\beta$ Pic disk demonstrated that a planet with a mass ratio between $10^{-2}$ to $10^{-5}$ on an inclined orbit can induce the necessary gravitational perturbations to create the warp structure of $\beta$ Pic's debris disk. \cite{dawson11} utilized $N$-body simulations to show that differential precession rates can lead to the formation of an inclined inner disk aligned with the planet, a flat outer disk, and a distinct warp at their interface. \cite{nesvold15} further integrated dynamical simulations with collision models to explain the warp and X-shaped pattern observed in the $\beta$ Pic disk. In these studies, the warp of the disk represents a transitional phase. According to \cite{brady23}, over long enough evolutionary timescales, the warp of the disk would be erased, and the planet would completely tilt the debris disk.

Most of these studies did not consider the role of the disk's self-gravity. In fact, the importance of the disk's self-gravity in the dynamics of debris disks has been increasingly recognized. \citet{poblete23} used $N$-body simulation to fully simulate the interaction between a planet and a debris disk composed of hundreds of particles. They found that considering the disk's self-gravity, a highly inclined planet would not lead to the destruction of the disk's structure, and the disk's morphology would be completely different from when self-gravity is not considered. \cite{sefilian24} pointed out through analysis that the precession speed of planetesimals caused by a massive debris disk's self-gravity could be an order of magnitude faster than that caused by planets, meaning the effect of the planet would be significantly weaker than predicted. Recently, \citet{sefilian25} systematically investigated the long-term vertical evolution of a self-gravitating debris disk under the perturbations from an inclined inner planet. Their work demonstrates that the disk-to-planet mass ratio dominates the system's dynamical state and significantly influences the propagation and persistence of the warp. Consequently, they highlight that neglecting the disk's self-gravity can hinder the reliability of inferring planetary parameters from observed disk structures.

Currently, to study the effects of self-gravity of debris disks, most studies use analytical methods \citep{sefilian21, sefilian23, sefilian24, sefilian25} or a smaller number of planetesimals \citep{poblete23}. Direct $N$-body simulations of debris disks containing a large number of interacting planetesimals remain rare, which give self-consistent results and better match the realistic conditions. Taking advantage of graphical processing unit (GPU) computing, we utilize the GENGA code \citep[GPU Gravitational ENcounters with Gpu Acceleration,][]{grimm14,grimm18,grimm22} to simulate the dynamics of debris disks composed of tens of thousands of particles. Our goal is to construct self-consistent dynamical evolution models of debris disks, and to investigate the warping structure induced by an internal Jupiter-mass planet misaligned with the disk's orbital plane.

We studied the warped debris disk around HD~110058 as an example. HD~110058 (HIP~61782) is an A0V star with an age of 17 Myr located at a distance of $130.08\pm0.53$~pc in the Scorpius-Centaurus (Sco-Cen) OB association. A nearly edge-on warped debris disk between 20 AU and 60 AU was resolved\citep{kasper15,esposito20,stasevic23, hom24, crotts24}. The debris disk shows a significant counter-clockwise S-shaped warp structure. These features strongly suggest the presence of an unseen non-coplanar planet orbiting inside the inner edge of the disk. 

In this work, using large-scale GPU $N$-body simulations, we find that the perturbation of an inner planet induces an approximate rigid-body precession of the self-gravitating disk, generating observable near-periodic warp structures. By systematically studying the relationship between planetary orbital parameters and debris disk warp configurations, we constrain the parameters of the unseen planet and the debris disk in the HD~110058 system. The structure of the paper is organized as follows. Section 2 provides a detailed description of our numerical simulation methodology, including the initial setup of the simulations. In section 3, the results of the simulation are studied in detail, and an empirical relationship between the warp structure and the system parameters is given. In Section 4, we discuss the morphology of the disk using the numerical simulation results and constrain the parameters of the system. In the final section we summarize the paper.

\section{$N$-body simulation and initial setup}
We performed batches of $N$-body simulations using the state-of-the-art $N$-body code GENGA \citep{grimm14,grimm18,grimm22}. 
$20,000$ mass-bearing particles are used in the simulation considering the gravitational forces between every pair of particles to model the influence of self-gravity of the debris disk. 

The initial setup of the simulation is based on the ground-based high-contrast observations of the HD~110058 system \citep{stasevic23, hom24, crotts24}. According to the fitting results from \citet{hom24}, HD~110058 has an inner boundary $<42$~AU, thus we set the disk's inner edge to 40~AU. For the disk's outer boundary, we set it to 80~AU, which is greater than the observed value ($\sim$ 60 AU) to avoid edge effects.
The particles have semi-major axes uniformly and randomly distributed. The total mass of the disk is set to be $100~M_\oplus$ as a standard case, which corresponds to the upper limit of a realistic estimate of the debris disk mass \citep{krivov21}. Additional cases with masses of $50~M_\oplus$ and $200~M_\oplus$ were also simulated. For simplicity, all the particles are assumed to have the same mass. 
In this setup, the surface density of the disk follows $\Sigma_d(r) = \Sigma_0 (r/r_\mathrm{out})^{-1}$. 
The disk is initially placed in a flat plane, and we choose this plane as the x-y plane of the coordinate system. We initialize the disk in a relatively cold state, with initial eccentricities following a Rayleigh distribution with an RMS (Root Mean Square) of 0.035, and initial inclinations following a normal distribution with RMS=1.42$^\circ$. The remaining orbital elements of planetesimals, namely longitude of periastron $\omega_\mathrm{par}$, longitude of ascending node $\Omega_\mathrm{par}$, and mean anomaly $M_\mathrm{par}$, were randomly and uniformly distributed between 0 and $2\pi$. 

\begin{figure*}[ht!]
\includegraphics[width=1\linewidth]{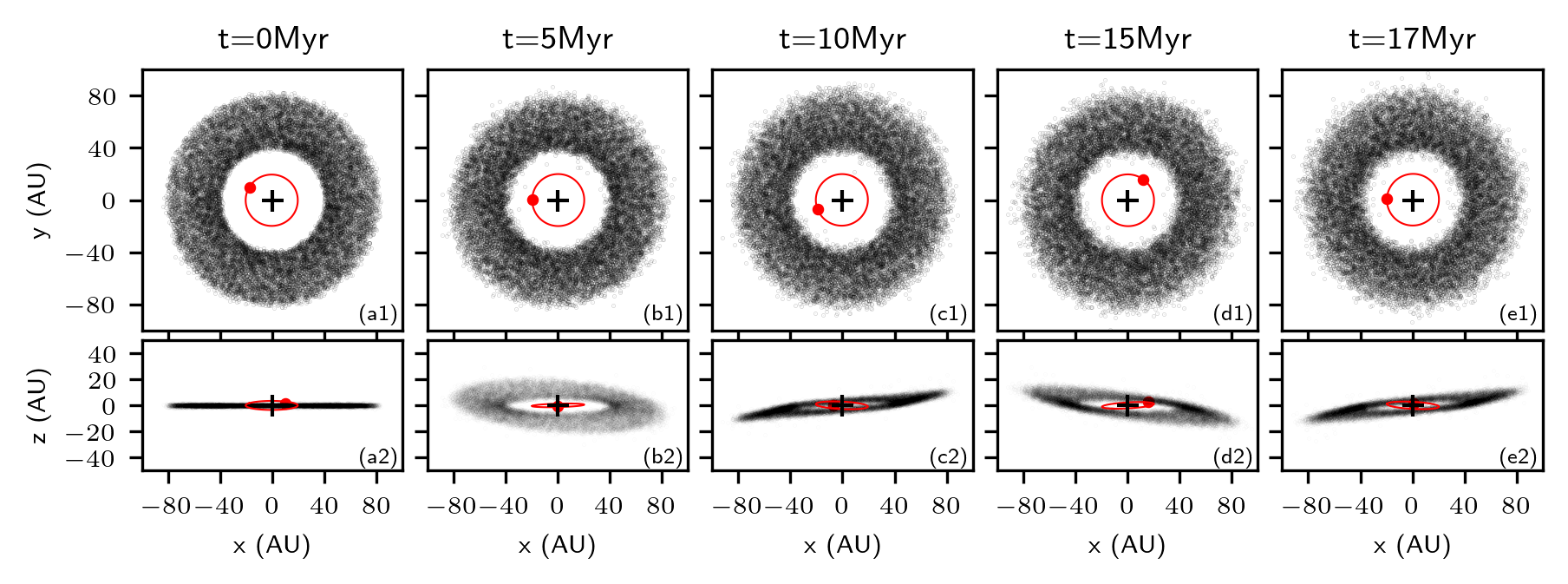}
\caption{Example snapshots of the debris disk for the standard case ($M_\mathrm{disk}=100M_\oplus$)at different epochs, including 0~Myr (initial setup), 5 Myr, 10 Myr, 15 Myr, and 17 Myr (the estimated age of the HD~110058). The black dots indicate the positions of the particles. The red curve and dot represent the instantaneous orbit and positions of the planet, and the black plus symbols indicate the central star. The upper panels (a1)-(e1) display the face-on perspectives of the disk, and the lower panels (a2)-(e2) display a nearly edge-on perspective.
\label{fig:fig1}}
\end{figure*}

Inside the disk, a planet with a typical mass of 1 Jupiter Mass ($M_\mathrm{J}$) moves in orbit with a semi-major axis of 20~AU and inclination of 10$^\circ$. Eccentricity of the planet is chosen to be 0.03, longitude of ascending node $\Omega_\mathrm{pl}=0$, and the remainder of the planet's orbital elements is randomly chosen between 0 and $2\pi$. Finally, the mass of the central star is set to be 2.1~$M_\odot$ \citep{chen14}. 
We simulate the system for 20~Myr, covering the 17~Myr age of the HD~110058 system, and the time step used was 1/20 of the planet's orbital period.

\change{In addition to the standard case ($M_\mathrm{disk}=100~M_\oplus$), we performed 14 additional simulations to explore the parameter space. In each set of simulations, we varied a single parameter while keeping the others fixed to the standard values. The variations include:
\begin{itemize}[noitemsep, topsep=0pt]
    \item Disk mass ($M_\mathrm{disk}$): $50~M_\oplus$ and $200~M_\oplus$;
    \item Disk inner radius ($r_\mathrm{in}$): 35~AU and 55~AU;
    \item Disk outer radius ($r_\mathrm{out}$): 75~AU and 85~AU;
    \item Planetary mass ($M_\mathrm{pl}$): $0.5~M_\mathrm{J}$ and $2.0~M_\mathrm{J}$;
    \item Planetary semi-major axis ($a_\mathrm{pl}$): 15~AU and 25~AU;
    \item Planetary inclination ($i_\mathrm{pl}$): $5^\circ$ and $15^\circ$;
    \item Number of particles ($N_\mathrm{par}$): 10,000 and 30,000.
\end{itemize}
}
\section{Results of $N$-body Simulation}
\subsection{Snapshots of the Disk}

Snapshots of the simulation of disk with mass of $100 M_\oplus$ at different epochs are shown in Figure \ref{fig:fig1}. It can be seen that the 20,000 particles (black dots) can clearly illustrate the structure of the disk. In the upper panels (a1) - (e1), the face-on perspectives of the disk at different epochs are shown. At the start of the simulation, the disk is initially in a relatively cool state, with sharp inner and outer edges. As the simulation progresses, the disk gradually becomes more diffuse due to planetesimal-planetesimal and planet-planetesimal gravitational interactions. In each of the lower panels, a nearly edge-on perspective of the disk is shown. It can be seen that the disk has an overall semi-rigid precession. Moreover, the disk also shows a notable warp structure in panels (c2) and (e2). Panels (e1) and (e2) display the simulation results at 17~Myr, corresponding to the estimated age of the HD~110058 system. At this stage, the simulated disk exhibits a pronounced warp, consistent with observations of the actual system. During the evolution, the orbit of planet also has a precession, as a result of gravitational perturbation by the disk.

\subsection{Evolution of Each Ring Forming the Disk}
To further investigate the warp structure of the disk, we divided the disk into 13 concentric rings\footnote{The disk was divided into 13 rings to facilitate the selection of 5 evenly spaced rings for plotting, including the innermost and outermost ones.}, which are composed of planetesimals within a certain range of distances from the star. Each ring has a width of 3.08~AU. To characterize the orientation of the $j$-th ring, we computed the average of the unit angular momentum vectors for all planetesimals within that ring using the expression:
\begin{equation}
    \bm{l}_j = \langle \bm{l}_{\text{par}, j} \rangle
    = \left\langle \frac{ \bm{r}_{\text{par}, j} \bm{\times} \bm{v}_{\text{par}, j} }
    {\lVert \bm{r}_{\text{par}, j} \bm{\times} \bm{v}_{\text{par}, j} \rVert} \right\rangle,
\end{equation}
where $\bm{l}_j$ represents the average unit angular momentum vector of the $j$th-ring, and it is perpendicular to the mid-plane of the ring, 
$\bm{l}_{\text{par}, j}$ is unit angular momentum vector of particle within the $j$-th ring, 
$\langle ~ \rangle$ indicates the averaging operation over all particles in the ring, and $\lVert ~ \rVert$ represents the magnitude of the vector.
The position vector and the velocity vector of each particle are represented by $\bm{r}_{\text{par}, j}$ and $\bm{v}_{\text{par}, j}$, respectively. 
Similarly, the unit angular momentum vector of the planet can be calculated as
\begin{equation}
\bm{l}_\text{pl} = \frac{\bm{r}_\text{pl} \bm{\times} \bm{v}_\text{pl}}
{\lVert \bm{r}_\text{pl} \bm{\times} \bm{v}_\text{pl} \rVert},
\end{equation}
where $\bm{r}_\text{pl}$ and $\bm{v}_\text{pl}$ are the position vector and the velocity vector of the planet, respectively.

\begin{figure*}[ht!]
\includegraphics[width=1\linewidth]{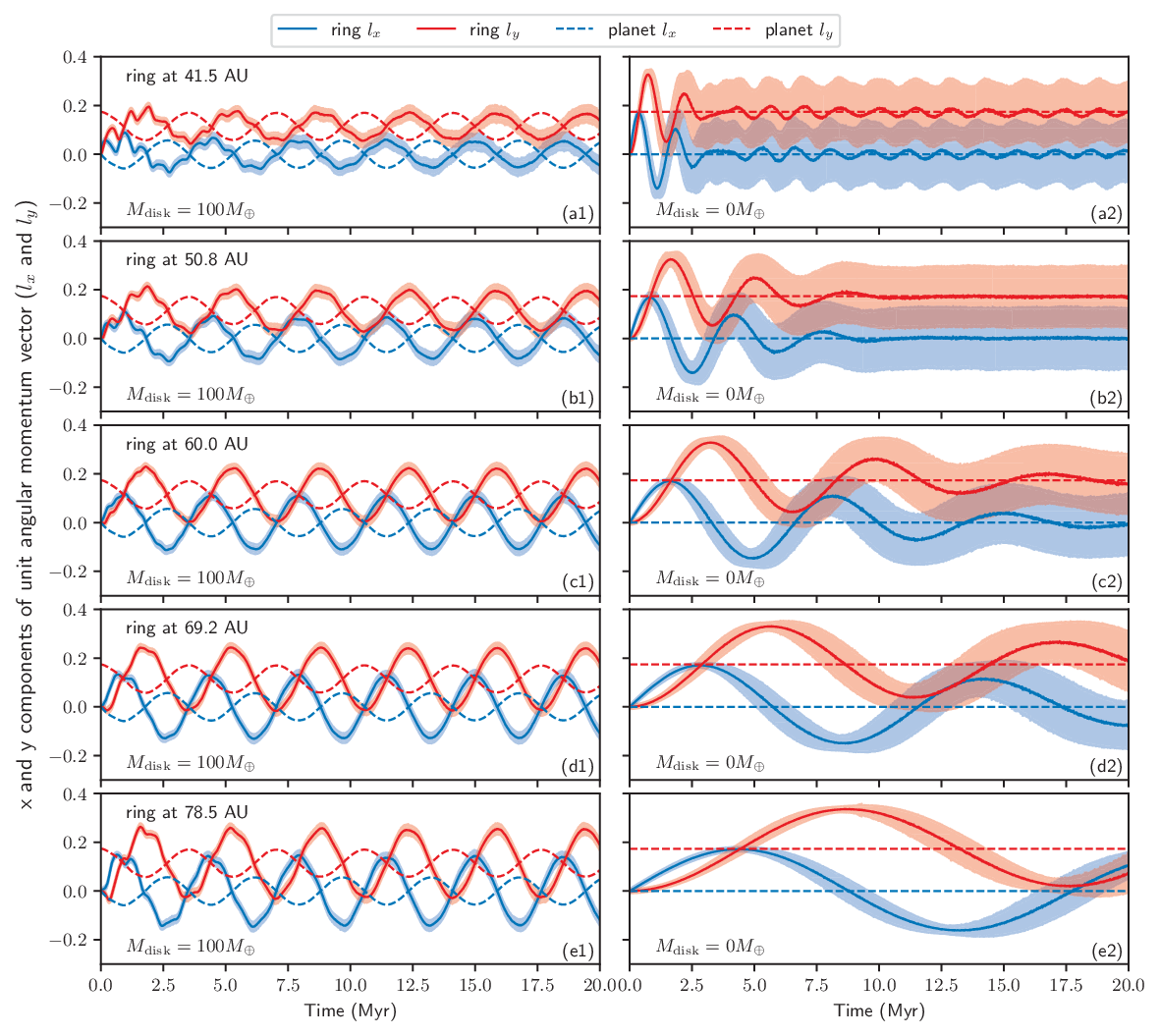}
\caption{The evolution of the unit angular momentum vectors of the concentric rings. Red and blue solid lines represent the $x$- and $y$-components of the ring's angular momentum. Red and blue dashed lines show those of the planet. The left five panels (a1-e1) display the case with disk mass of $100~M_\oplus$. The right five panels (a2-e2) show the case with a massless disk formed by test particles.
\label{fig:fig2_lxly}}
\end{figure*}

In Figure \ref{fig:fig2_lxly}, we present the evolution of $x-$ and $y-$ components of the unit angular momentum vectors of the concentric rings ($l_{x,j}$, $l_{y,j}$; red and blue solid lines) and those of the planetary orbit ($l_{x,\mathrm{pl}}$, $l_{y,\mathrm{pl}}$, red and blue dashed lines). The shaded regions indicate the standard deviation of $\bm{l}_\mathrm{par}$. Among the 13 concentric rings, five selected rings are displayed in panels (a1-e1), with their centers at radii of 41.5~AU (innermost), 50.8~AU, 60.0~AU, 69.2~AU, and 78.5~AU (outermost), respectively. 
From the figure, we can observe the following features:
\begin{itemize}[noitemsep, resume, topsep=0pt, label=--]
\item The $l_{x,j}$ and $l_{y,j}$ components of all rings show periodic oscillations with two main periods. The dominant period is about 3.5~Myr, and another shorter period is about 0.7~Myr.
All the rings have the same precession periods and phases regardless of radial distance, suggesting the entire disk undergoes rigid-body-like precession.
\item The planetary angular momentum follows a single-frequency oscillation, with the same period as the rings' primary component, which means the rings and planet share a common precession. The angular momentum components $l_x$ and $l_y$ of both the planet and the rings oscillate around the same midlines, which correspond to the total angular momentum vector of the system.
\item The oscillation amplitude of the primary component of the rings increases with heliocentric distance. This radial dependence confirms a warped geometry of the disk. 
\item The secondary oscillation components of the rings undergo gradual damping with a timescale $\sim 10$~Myr. 

\item The spreads of $l_{x, j}$ and $l_{y, j}$ grow during evolution (represented by the blue and red shaded regions in Figure \ref{fig:fig2_lxly} respectively), indicating the increase of the disk's vertical scale height.
\end{itemize}
In the right-hand panels (a2-e2) of Figure \ref{fig:fig2_lxly}, the additional simulation results of a non-self-gravitating disk of massless test particles are plotted, which have also been analyzed in detail in \citet{dawson11}, \citet{nesvold15}, and \citet{brady23}. In the absence of inter-ring gravitational interactions, each ring undergoes independent differential precession governed by the secular torque from the planet, with a secular timescale $\propto r_j^{7/2}$ \citep{brady23}. This radial dependence drives asynchronous phase evolution: inner rings rapidly precess leading to complete phase mixing (e.g. $3~\mathrm{Myr}$ for the ring at 41.5~AU) and vertical thickening ($h_j \approx 2r_j \cdot i_0$), while outer rings do not mix as much due to slower precession. Over time, phase mixing propagates outward, resulting in an outward propagation of the warp's position. 

\subsection{Evolution of the Warp Structure}
\begin{figure}[h!]
\includegraphics[width=1\linewidth]{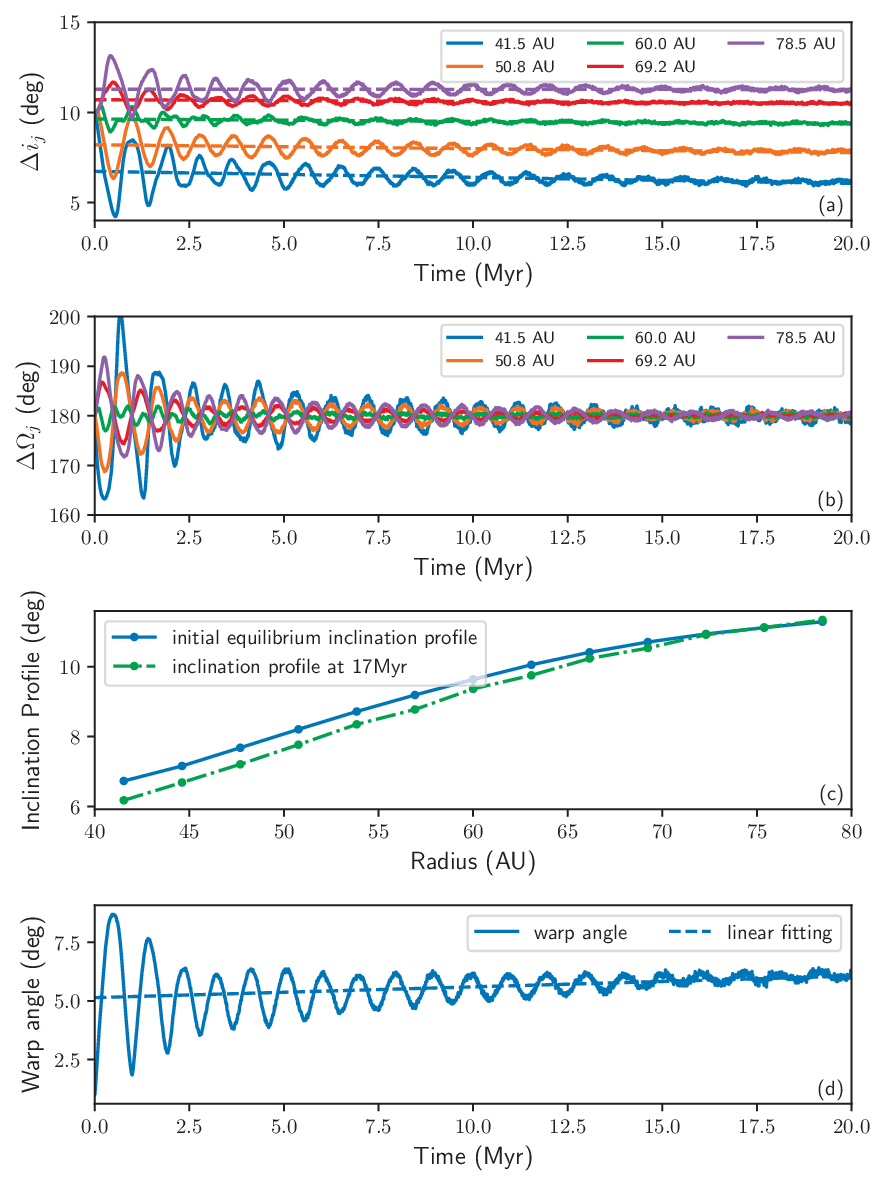}
\caption{Evolution of relative inclinations of the rings. Panel (a): Solid lines show the evolution of relative inclination of the ring with respect to the planetary orbital plane. Dashed lines indicate the linear fitting results. \change{Panel (b): Evolution of the relative longitude of the ascending node, $\Delta \Omega_j$, of the rings with respect to the planetary orbital plane.}. Panel (c): Green dot-dashed line with points indicates the inclination profile of the disk at 17~Myr, Blue line with points shows the initial equilibrium inclination profile from linear fitting of $N$-body simulation results. Panel (d): Solid line displays the evolution of the warp angle (relative angle between the innermost and outermost rings. Dashed line shows linear fitting result).}
\label{fig:warp_evo}
\end{figure}

Since the rings and planet share a common precession, we calculate their relative inclination to remove the coherent precession component, focusing on the evolution of the finer warp structures of the disk. The relative inclination of the $j$-th ring is defined as
\begin{equation}
\Delta i_j = \arccos(\bm{l}_j \bm{\cdot} \bm{l}_\mathrm{pl}).
\end{equation}
Figure \ref{fig:warp_evo}a illustrates the evolution of $\Delta i_j$ for the five rings (41.5, 50.8, 60.0, 69.2, 78.5~AU), indicated by blue, orange, green, red and purple solid lines, respectively. All rings undergo damped oscillations around certain equilibrium values: inner rings (blue/orange) oscillate near $6^\circ$-$8^\circ$, while outer rings (red/purple) oscillate near $10^\circ$-$11^\circ$. Dashed lines in the figure show the results of the linear fitting of $\Delta i_j$ versus time, revealing the long-term trends of the equilibrium angles. The equilibrium angles of inner rings show $\sim0.5^{\circ} / \mathrm{Myr}$ downward drift, and it is $\sim0.2^{\circ} / \mathrm{Myr}$ for outer rings.
The differential relative inclinations indicate a warped disk configuration: the inner region lies closer to the planet orbital plane while the outer region is further. 

\change{Figure \ref{fig:warp_evo}b displays the evolution of the relative longitude of ascending node ($\Delta \Omega_j$) for the rings. The nodes of all rings oscillate around $180^\circ$. In the early stage of evolution ($t < 2.5$~Myr), the phase difference between the innermost and outermost rings reaches up to $\sim 30^\circ$. This indicates a significant twist in the disk structure. As the evolution progresses, this twist diminishes, and the rings gradually align, eventually exhibiting a global, coherent bending.}

Using linear fitting results of $\Delta i_j$, (dashed lines in panel a), we obtain the profile of the equilibrium inclination of the rings. The initial equilibrium inclination profile (t=0) is plotted in Figure \ref{fig:warp_evo}c (blue solid line with points). Additionally, the disk's inclination profile at 17~Myr (the age of HD~110058) is shown as a green dot-dashed line with data points. The oscillation amplitude of the warp is almost fully dissipated by 17~Myr, and the disk will maintain an equilibrium warp configuration while exhibiting nodal precession at the same rate as that of the planet. Although the inclinations between the rings and the planet are damping, the warping degree of the disk increases. This is because the inner rings damp faster than the outer ones.

\begin{figure*}[htb!]
\includegraphics[width=1\linewidth]{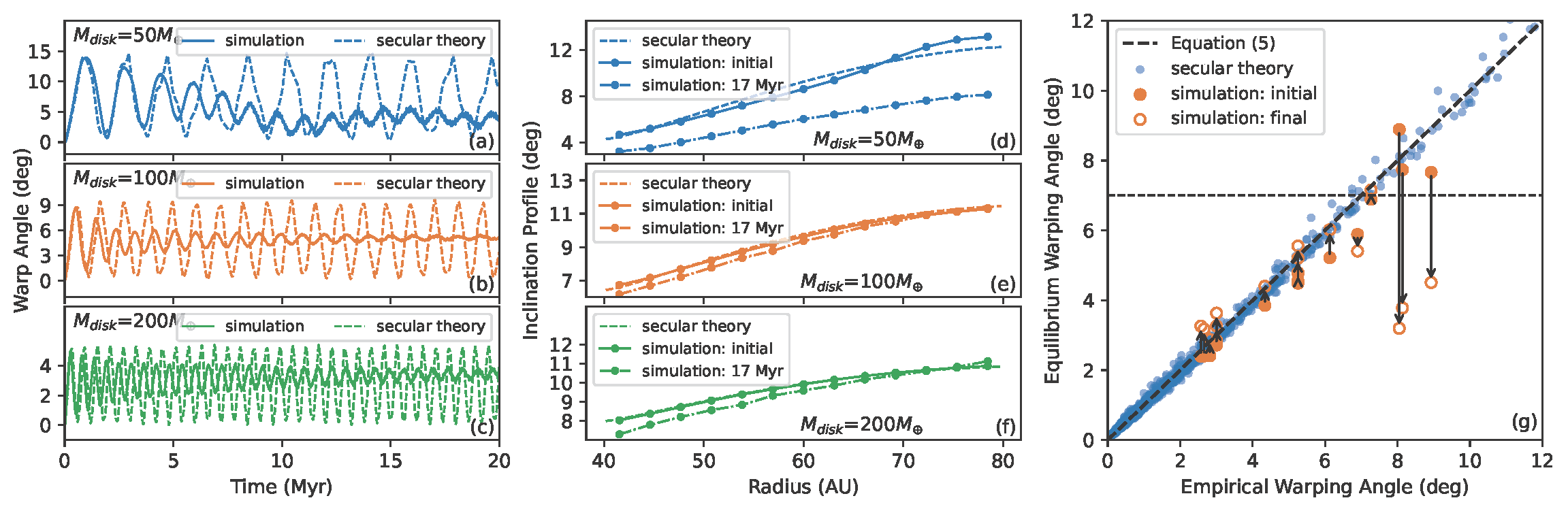}
\caption{Comparison between N-body simulation and theoretical analysis. Panels (a), (b), and (c): Evolution of the warp angle for disk masses of $50~M_\oplus$, $100~M_\oplus$, and $200 ~M_\oplus$, respectively. Solid lines represent the simulation results, while dashed lines show those from Laplace-Lagrange secular theory. Panels (d), (e), and (f): Initial equilibrium inclination profile for the same three disk masses. Solid lines are from simulations, and dashed lines are from the analytical model. \change{The dot-dashed lines with points indicate the inclination profile of the disk at 17~Myr.} Panel (g): The relationship between the equilibrium warp angle and system parameters. The x-axis shows the warp angle predicted by Equation \ref{eq:trends}. Blue dots represent results from the Laplace-Lagrange theory. \change{The initial warp angles and those at 17 Myr of $N$-body simulation are presented using orange dots and circles.}}
\label{fig:disk_mass}
\end{figure*}

To quantify the global warping, we define the warp angle $\Delta i_{\rm warp}$ as the relative inclination between the innermost ring ($j=1$) and outermost ring ($j=13$),
\begin{equation}
\Delta i_{\rm warp} =\Delta i_{13} - \Delta i_{1}
\label{eq:warp_angle}
\end{equation}
Figure \ref{fig:warp_evo}(d) shows $\Delta i_{\rm warp}$ evolution with time. The initially flat disk ($\Delta i_{\rm warp}=0^\circ$) rapidly warps to $\sim 9^{\circ}$ within 0.5~Myr. Subsequently, the warp angle oscillates around 5$^{\circ}$-6$^{\circ}$ with a period of about 1~Myr. Oscillation amplitudes damp over time, approaching a near-equilibrium state with $\Delta i_{\rm warp} \sim 6^{\circ}$. The linear fitting is shown by blue dashed line, indicating a positive trend due to the faster damping rate of the inner rings. 

\subsection{Analytic Study of the Warp Structure}

Using analytic methods, similar equilibrium warp configurations in self-gravitating disks have been found \citep{ulubay-siddiki09, zanazzi18, batygin18, melton21}. In these studies, the disk is typically approximated as a razor-thin plane. It is then divided into a series of concentric narrow rings (wires) to approximate the dynamical evolution of the disk. The Laplace-Lagrange secular perturbation equations are widely applied in related research \citep{batygin11, batygin18, melton21}. Following \citet{batygin11}, but adapting the equation for an internal perturber as described by \citet{murray99}, we theoretically model the disk with 100 wires. To cross-verify the simulation and theoretical analysis, we performed simulations and analysis of three cases with different disk masses of 50 $M_\oplus$, 100 $M_\oplus$, and 200 $M_\oplus$, respectively. In these simulations, all other parameters were consistent with the standard model described in the previous section. In all three cases, the total number of particles is 20,000, with individual particle masses adjusted according to the disk's mass.

In panels (a), (b), and (c) of Figure \ref{fig:disk_mass}, the evolutions of the warp angles from both simulation (solid lines) and analytic theory (dashed line) are displayed. Starting from a flat configuration (warp angle $=0^\circ$), the disk's warp angle oscillates around an equilibrium value. Lower-mass disks exhibit larger warp angles and longer oscillation periods. 
For the first oscillation period, $N$-body simulation and theoretical analysis agree with each other well. However, they diverge over longer timescales. The oscillation amplitude damps over time, which is not observed in the theoretical result. This damping is likely due to dissipative effects within the disk, such as dynamical friction caused by planetesimal scattering, which are not included in the secular theory. 
\change{According to \citet{melton21}, the evolution of a self-gravitating disk involves the superposition of multiple oscillation modes, each characterized by specific eigenvalues (precession frequencies) and eigenvectors (warp profiles). For the $100~M_\oplus$ disk, the secular theory yields the first two non-zero eigenvalues as $92^\circ/\mathrm{Myr}$ (corresponding to a period of $3.90~\mathrm{Myr}$) and $420^\circ/\mathrm{Myr}$ ($0.86~\mathrm{Myr}$). These match the dominant periods derived from our $N$-body simulations ($3.5~\mathrm{Myr}$ and $0.7~\mathrm{Myr}$) well. During the evolution, higher-order oscillation modes damp out rapidly due to dissipation, eventually leaving only the fundamental mode, which corresponds to the equilibrium warp profile.} 

Panels (d), (e), and (f) compare the initial equilibrium inclination profiles. The theoretical and simulation results are also in good agreement, particularly for massive disks of 100 $M_\oplus$ and 200 $M_\oplus$. For the 50 $M_\oplus$ disk, differences in the profile shape can be observed. This is likely because the dissipation occurs rapidly within the first 10~Myr, and the equilibrium profile derived from simulation using linear fitting may be less precise. However, the resulting overall warp angle remains consistent with the theory with $\sim 10\%$ difference. \change{We also plot the inclination profile at the system's current age (17~Myr) using a dot-dashed line. For the low-mass case ($50~M_\oplus$), the profile at this late stage deviates significantly from the initial equilibrium state. This divergence is likely because the planetary perturbation is too large compared with the self-gravity of the less massive disk, potentially leading to the disruption or breaking of the disk structure over long timescales. We will investigate this mechanism further in future work.}

Conducting a large-scale parameter study of how disk and planet properties influence the warp structure using $N$-body simulations would be extremely time-consuming. Since we have verified that the results from the Laplace-Lagrange secular theory closely match those of $N$-body simulations, the more efficient analytical method is used to explore the parameter space. We generate 1000 random samples within the following parameter space:
\begin{itemize}[noitemsep, topsep=0pt, label=--]
    \item Disk mass ($M_{\rm disk}$): [50~$M_\oplus$, 200~$M_\oplus$],
    \item Disk inner radius ($r_{\rm in}$): [30~AU, 50~AU],
    \item Disk outer radius ($r_{\rm out}$): [$r_{\rm in}$+20~AU, $r_{\rm in}$+40~AU], 
    \item Planetary mass ($M_{\rm pl}$): [0.5~$M_J$, 1.5~$M_J$],
    \item Planetary semi-major axis ($a_{\rm pl}$): [10~AU, 20~AU],
    \item Planetary inclination ($i_{\rm pl}$): [5$^\circ$, 15$^\circ$].
\end{itemize}
\change{Based on these results, we fit the equilibrium warp angle using a functional form dependent on the system parameters:
\begin{equation}
\begin{aligned}
i_{\rm warp} =  i_{\rm pl} \left\{0.83 + 0.32\left[\frac{M_{\rm pl}(1 - \sin i_{\rm pl})}{M_{\rm disk}}\right]^{-0.75}  \right.\\
\left.\times\left(\frac{a_\mathrm{pl}}{r_{\rm in}}\right)^{-2.43}  \left(\frac{r_{\rm out}-r_{\rm in}}{r_{\rm in}}\right)^{-1.23}\right\}^{-1},
\label{eq:trends}
\end{aligned}
\end{equation}}
\change{The coefficients are fitted using SciPy's curve\_fit function. We examine the quality of this fit with respect to each variable in Appendix A, finding good agreement across the explored parameter space.} Note that $M_{\rm pl}(1 - \sin i_{\rm pl})$ in the expression is due to the projection of planetary mass onto the disk's reference plane \citep[see][]{batygin11}. 


Panel (g) of Figure \ref{fig:disk_mass} validates this relationship. The warp angles calculated from Laplace-Lagrange secular theory (blue dots) show a strong correlation with the predictions of Equation \ref{eq:trends}. \change{The initial equilibrium warp angles of a total of 15 runs of $N$-body simulations (orange dots) also align well with this theoretical trend, confirming the equation's validity. We also plot the warp angles at 17~Myr using open circles. As shown in the Panel (g), for systems with an initial warp angle exceeding $\sim 7^\circ$, the warp angle decays rapidly to $3^\circ-4^\circ$. This suggests that maintaining a warp larger than $7^\circ$ over 17 Myr is dynamically difficult for initially flat disks, placing an implicit upper limit on the sustainable warp amplitude.} 

\section{Discussion}
\subsection{Morphology of the Simulated Disk}
To compare our model with observations, we generate synthetic images from the $N$-body simulation results. We use the simulation snapshot of $100~M_\oplus$ disk at 17~Myr to study the disk morphology. Following the method of \citet{poblete23}, each particle's orbit is populated with 10 dust grains. The intensity of starlight scattered by each grain is then calculated following the method of \citet{crotts24}. The resulting scattered flux is projected onto a 100$\times$100 pixel grid with a pixel scale of 0.97~AU/pixel to match the resolution of SPHERE/IFU. Finally, the synthetic image is convolved with a Gaussian kernel with a full width at half maximum (FWHM) of 6.5 AU, corresponding to the SPHERE H-band point-spread function (PSF). 
\change{Since the observed morphology of a warped disk depends on the viewing angle relative to the warp's line of nodes, we tested various observer orientations.
When the viewing angle is perpendicular to the line of nodes, the observed warp is maximized. As the viewing angle shifts, the projected warp decreases roughly following a cosine relationship, consistent with \citet{mouillet97}. Since our simulated intrinsic warp is $\sim 6^\circ$, a $30^\circ$ offset naturally reduces the projected warp to match the observational $\sim 5^\circ$ warping.}

\begin{figure*}[ht!]
\includegraphics[width=1\linewidth]{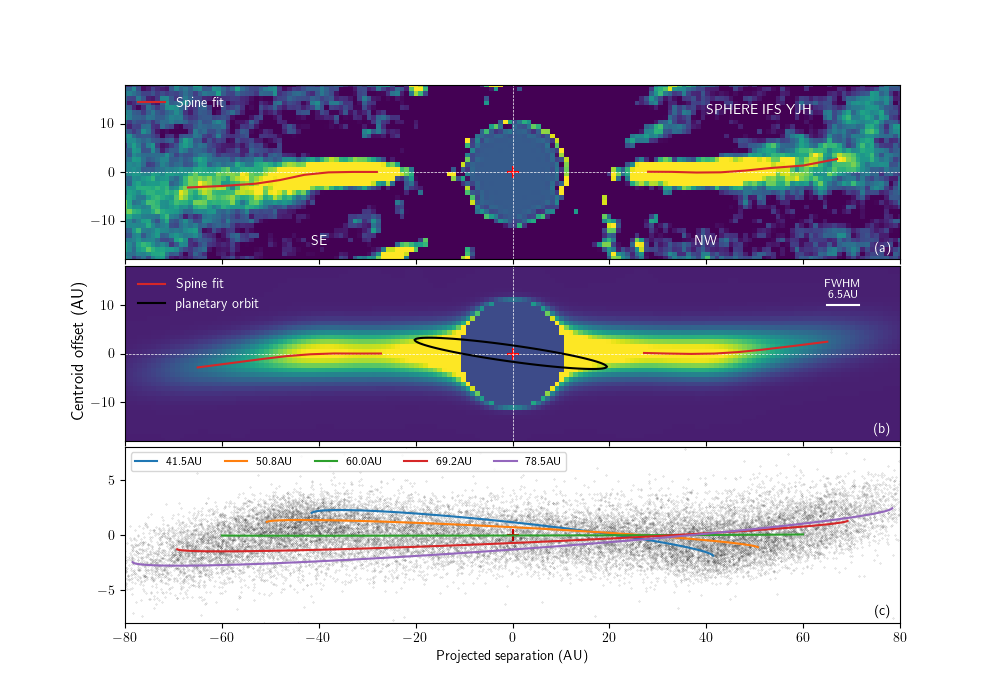}
\caption{Comparison between observed and simulated images of the HD 110058 debris disk. Panel (a): The observed image from \citet{stasevic23}, identical to the lowest panel of their Figure 5. The red solid line is the fitted spine of the disk. Panel (b): The synthetic image from our $N$-body simulation. Panel (c): Spatial distribution of particles in the $N$-body simulation. The colored solid lines mark the five rings in Figure \ref{fig:fig2_lxly} and \ref{fig:warp_evo}. Due to the forward scattering of the dust grains, we only plot the front side of the rings. 
\label{fig:morphology}}
\end{figure*}

Figure \ref{fig:morphology} compares our synthetic image with the observed images of the HD 110058 debris disk. Panel (a) shows the combined YJH-band data of HD110058 from \citet{stasevic23}. The red solid lines trace the disk's spines of the two limbs, fitted using the same method as \citet{stasevic23}. The image is rotated clockwise by 66.2$^\circ$ to align the inner disk ($<40$ AU) with the $x$-axis, where the right and left limbs correspond to the Northwest (NW) and Southeast (SE) sides. The distinct warp structure at $r_\mathrm{wrap} \sim 40$ AU is clearly observed.   

Panel (b) shows the synthetic image from our $N$-body simulation. The simulated disk exhibits a similar structure to the observations, including the warp location ($r_{\rm wrap} \approx 40~\mathrm{AU}$) and the $\sim 5^\circ$ warp difference between the inner and outer disks. For the NW spine, the downward-then-upward bending feature is well reproduced.

Panel (c) reveals the underlying particle distribution that creates this morphology, where the colored lines trace the orbital planes of five distinct rings (see Figures \ref{fig:fig2_lxly} and \ref{fig:warp_evo}). The apparent warp is a projection effect arising from the misalignment of these rings. In the inner region ($<40$~AU), the projections of the rings overlap, creating the appearance of a flat disk. In the outer region, the projected envelope of these tilted rings creates the observed asymmetry: a flat upper edge and a curved lower edge on the NW side and vice versa for the SE side. Our model also naturally explains the observed brightness and thickness asymmetry between the two sides of the disk. Due to the forward scattering property of the dust grains, the side of the disk with nodal lines toward the observer (the NW side in our model) appears thinner and brighter. This behavior is qualitatively consistent with the observed image, further strengthening the validity of our warped disk model.

\subsection{Constraints on Planetary and Disk Parameters}
\label{sec:constraints}
\begin{figure}[ht!]
\includegraphics[width=1\linewidth]{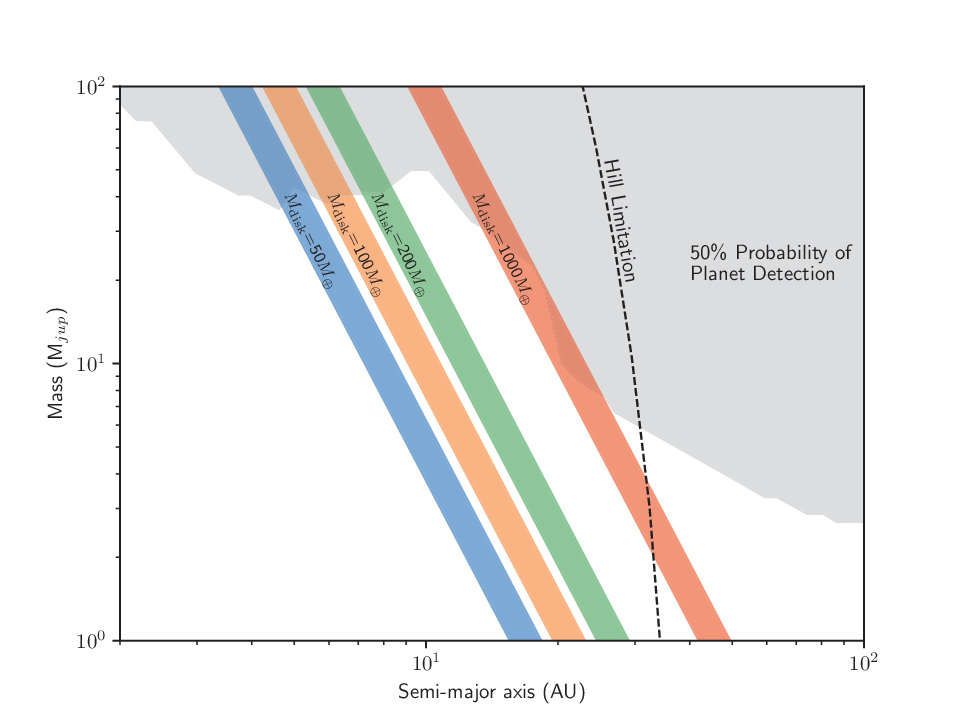}
\caption{The constraints on a potential planet's mass versus its semi-major axis. The gray shaded region indicates the 50\% detection probability limit from current observations \citep{stasevic23}; since no planet has been detected, its parameters are presumed to lie outside this region. The black dashed line represents the dynamical stability limit at three times the planet's Hill radius, requiring the planet to be located to the left of this line to avoid disrupting the inner edge of the disk. 
\change{The colored shaded regions represent the parameter space (planetary mass and semi-major axis) capable of producing the observed warp for disks of different masses.}}
\label{fig:pl_limit}
\end{figure}

By applying this relationship of Equation~\ref{eq:trends}, we can place constraints on both the debris disk mass and properties of an unseen planet within the HD~110058 system.
Figure \ref{fig:pl_limit} displays the constrained parameters in the parameter space of planetary mass versus semi-major axis. For a given initial inclination of $10^\circ$, the colored regions plot the parameters of the planet for various disk masses (e.g., 50, 100, 200, and 1000~$M_\oplus$ ) that produce the observed warp. \change{A planet that is not massive enough or located too far from the disk cannot generate the observed $5^\circ$ warp. However, our $N$-body simulations indicate that warps larger than approximately $7^\circ$ decay rapidly during the 17~Myr evolution. This provides an upper bound on the planetary perturbations. Therefore, we define the parameter space producing a warp between $5^\circ$ and $7^\circ$ as the range consistent with observations.} 

The gray shaded region indicates the 50\% detection probability zone from IRDIS/IFS direct imaging and Gaia astrometry \citep[from Figure 15 of][]{stasevic23}. Due to the non-detection of the planet, it is highly probable that the planet is not in this region. The dashed line marks the 3$\times$ Hill radius stability boundary, beyond which the inner edge of the disk ($\sim40$~AU) will be disrupted. Based on these constraints, we find that a low-to-moderate disk mass ($50-200~M_\oplus$) allows for a wide range of planetary parameters, from approximately one to several tens of Jupiter masses, to create the observed structure. Conversely, a significantly more massive disk of 1000 $M_\oplus$ requires a more massive perturber due to its strong self-gravity, which will reduce the possible parameter space. We therefore infer that a disk mass exceeding 1000 $M_\oplus$ is unlikely.

\subsection{Debris Disk Mass Problem}
As one of the most fundamental parameters, the total mass of a debris disk is still highly uncertain. As discussed in \citet{krivov21}, the disk's mass is thought to be dominated by unobservable kilometer-sized planetesimals, while observations can only detect micron-to-millimeter-sized dust and grains. The total mass must be inferred by extrapolating from (sub)millimeter observations, based on theoretical models of a dust-producing collisional cascade. Current observations have led to inferences of total debris disk masses as high as 1,000~$M_\oplus$ to~10,000 $M_\oplus$. However, according to planet formation models, a protoplanetary disk can support at most 100~$M_\oplus$ to 1,000~$M_\oplus$ of solids. This contradiction is known as the ``debris disk mass problem'' \citep{krivov21}. By considering the disk's self-gravity, it is possible to use the mutual dynamical interactions between the planet and the disk to constrain the disk's mass. Although parameters such as the semi-major axis, the planet-to-disk mass ratio, and the initial inclination are degenerate, and various combinations can produce a similarly warped disk, our model still allows us to infer that a disk with mass $\gtrsim$ 1,000 $M_\oplus$ is unlikely to form the observed warp structure. This conclusion supports the mass limits derived from protoplanetary disk theory. The future detection of a planet within the system would allow us to further constrain the properties of the disk.

\change{\citet{sefilian25} also analyzed the HD~110058 system and suggested a disk mass upper limit of $\sim 300~M_\oplus$. This estimate is lower than our constraint of $< 1,000~M_\oplus$. The discrepancy primarily arises because their model focused on the secular back-reaction of the disk on the planet but did not fully capture the internal self-gravity that allows the disk to maintain coherence. In our model, self-gravity acts as a restoring force, allowing the disk to precess as a semi-rigid body rather than breaking or phase-mixing incoherently. This cohesive effect enables more massive disks to sustain the observed warp geometry without disrupting, thereby supporting a higher mass upper limit.}

\subsection{Effects of Particle Resolution}
\change{In our standard simulations, we used $N=20,000$ particles. To validate the robustness of our results, we performed comparative simulations with $N=10,000$ (low resolution) and $N=30,000$ (high resolution), keeping the total disk mass constant at $100~M_\oplus$. The results (shown in Appendix A) demonstrate that varying the particle number has a negligible effect on the initial equilibrium warp angle. However, the particle number does influence the damping rate. Simulations with fewer particles exhibit faster dissipation due to increased granularity and scattering. At the end of the evolution (17 Myr), the low-resolution cases show a smaller relative inclination between the disk and planet, but a slightly larger internal warp angle due to differential damping between the inner and outer disk. Notably, this trend is consistent with the low-resolution (200 particles) model of \citet{poblete23}, where the disk became fully coplanar with the planet within 3~Myr. Our results confirm that while resolution affects the timescale of relaxation, the formation and morphology of the warp are robust.}

\section{Conclusion and Future Work}
In this work, we performed large-scale $N$-body simulations of the HD~110058 system to investigate the formation of its warped debris disk under the influence of an unseen, non-coplanar planet and the disk's self-gravity. Our key findings are as follows:
\begin{enumerate}[noitemsep, resume, topsep=0pt]
    \item The disk's self-gravity helps maintain the overall structure of the disk; as a result, the debris disk precesses like a semi-rigid body around the total angular momentum axis of the planet-disk system. Considering the self-gravity, planetary perturbations can establish a global warp across the entire disk in less than 0.5 Myr, which is different from the outward-propagating warps in massless disks.
    \item The degree of the warp of the disk undergoes a damped oscillation, and the disk will eventually settle into an equilibrium configuration. Over time, the relative inclination between the disk and the planet gradually decreases.
    \item The equilibrium warp state is well-described by Laplace-Lagrange secular theory. Using secular perturbation theory, we found a relationship (Equation \ref{eq:trends}) linking the equilibrium warp angle to the parameters of the planet and the disk. Using the relationship, we constrain the mass of the debris disk of HD 110058 to be $\lesssim 1,000~M_\oplus$ to form the observed warp structure. 
    \item Our synthetic scattered-light images of the debris disk, based on the $N$-body simulation results, successfully replicate the observed morphology of HD~110058. 
    The simulated disk image has a spine that closely matches the observations.
\end{enumerate}
In this study, we have focused primarily on the dynamical evolution of the ``parent disk'' composed of planetesimals, and assumed that dust particles follow the same orbits as the planetesimals. This assumption does not account for the evolution of dust, which is dominated by non-gravitational forces such as radiation pressure or stellar winds \citep{wyatt99}. Consequently, the dust might have a spatial distribution different from that of the planetesimals. \change{Recent observations suggest that HD~110058 is a gas-rich system \citep{hales22}. However the effect of gass was not included in our purely gravitational $N$-body simulations. The presence of gas can introduce aerodynamic drag on the dust and smaller planetesimals, and damp the inclinations and eccentricities.
We will carry out a detailed study of dust dynamics and effect of the gas in future work.}

We expect that the James Webb Space Telescope will provide higher signal-to-noise images of this system, potentially even detecting the perturbing planet, which would greatly refine the estimates of key parameters such as the disk mass. Furthermore, several high-resolution direct imaging instruments currently under development such as the Roman Coronagraph Instrument (CGI) \citep{kasdin20} and the Cool Planet Imaging Coronagraph (CPI-C) \citep{dou25} on the China Space Station Survey Telescope (CSST) \citep{csst25}, are expected to detect more warped debris disks, allowing for a deeper understanding of the dynamics of such structures.

\section{acknowledgments}
\change{We thank the anonymous referee’s valuable and thoughtful comments.} This work is supported by the National Natural Science Foundation of China (grant Nos. 12473076, U2031210, 11827804), the Natural Science Foundation of Jiangsu Province (grant No. BK20221563), the Foreign Expert Project (grant No. S20240145), as well as the science research grants CMS-CSST-2021-A11, CMS-CSST-2021-B04, CMS-CSST-2025-A17, CMS-CSST-2025-A18, and CMS-CSST-2025-A19 from the China Manned Space Project.

%

\vspace{5mm}


\software{GENGA \citep{grimm14,grimm18,grimm22}, NumPy \citep{harris20}, \change{SciPy} \citep{scipy2020}, Astropy \citep{astropycollaboration13, astropycollaboration18}, Matplotlib \citep{hunter07}}



\appendix
\section{Parameter study of the disk warp structure.}
\begin{figure*}[ht!]
\includegraphics[width=1\linewidth]{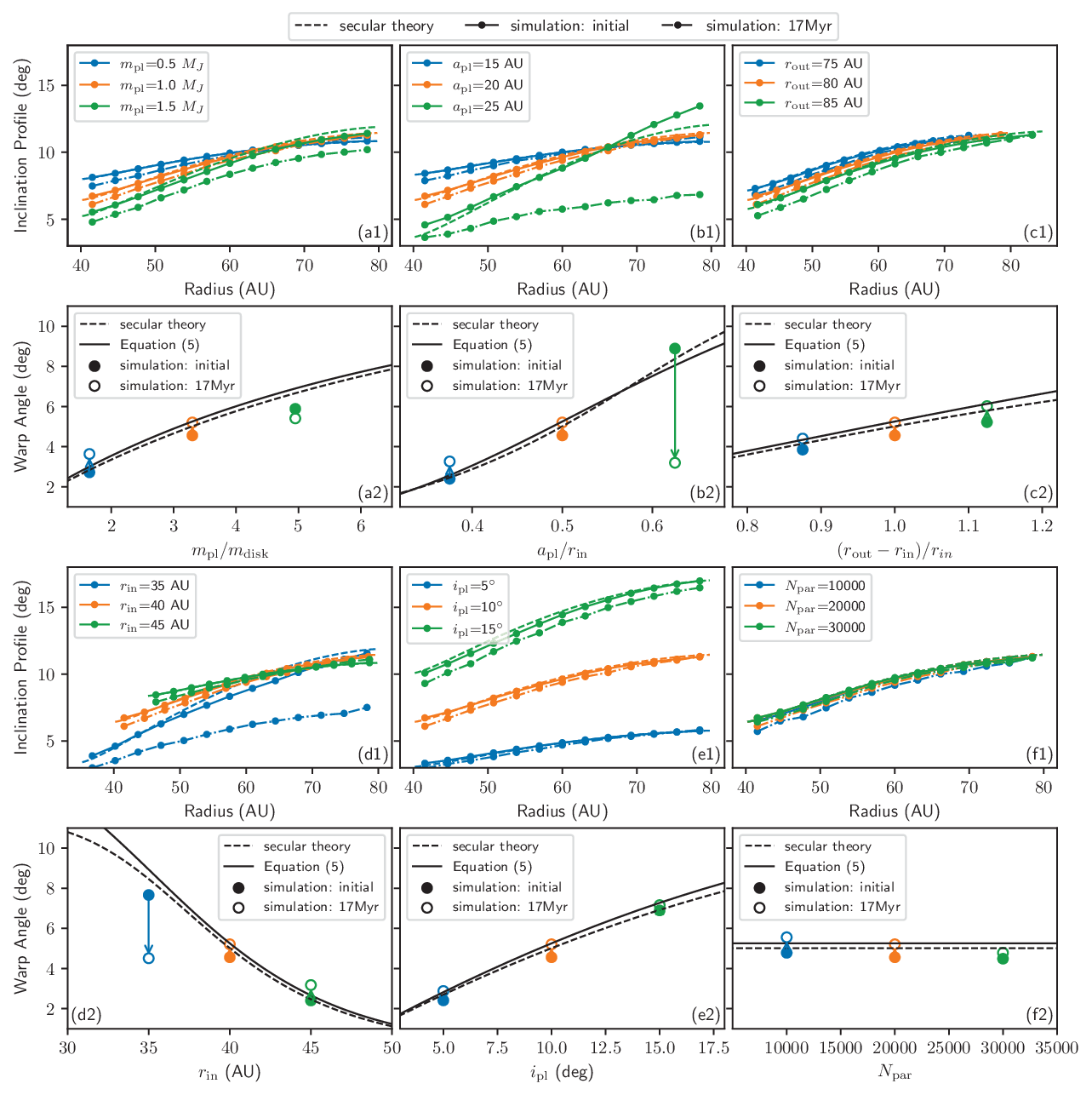}
\caption{Parameter study of the disk warp structure. 
The panels (a1) - (f1) display inclination profiles for variations in planetary mass ($M_\mathrm{pl}$), semi-major axis ($a_\mathrm{pl}$), disk outer radius ($r_\mathrm{out}$), disk inner radius ($r_\mathrm{in}$), planetary inclination ($i_\mathrm{pl}$) and particle number ($N_\mathrm{par}$). 
Line styles distinguish the initial equilibrium from $N$-body simulations (solid line with points), the state at 17~Myr (dot-dashed line with points), and secular theory predictions (dashed line). 
The panels (a2) - (f2) show the dependence of the warp angle on relevant parameters. 
Black solid and dashed lines represent the empirical fit (Equation~\ref{eq:trends}) and secular theory results, respectively. 
Colored markers denote $N$-body simulation results at the initial (solid circles) and at 17~Myr (open circles), with arrows indicating the evolutionary trend due to dissipation.}
\label{fig:parameter_study}
\end{figure*}

\change{We conducted a comprehensive parameter study involving 15 sets of $N$-body simulations. Figure \ref{fig:parameter_study} summarizes the results that are not detailed in the main text. Panels (a1)--(f1) compare the inclination profiles derived from $N$-body simulations with secular theory predictions while varying planetary mass ($M_\mathrm{pl}$), semi-major axis ($a_\mathrm{pl}$), inclination ($i_\mathrm{pl}$), disk radii ($r_\mathrm{in}$, $r_\mathrm{out}$), and particle count ($N_\mathrm{par}$). In all configurations, the initial equilibrium profiles from simulations (solid lines with points) show remarkable agreement with the theoretical models (dashed lines).}

\change{Panels (a2)--(f2) validate our empirical fit (Equation \ref{eq:trends}). In each panel, we vary a single parameter while holding others constant, comparing the predictions from Laplace-Lagrange secular theory (dashed lines) with our fitted model (solid lines). The two methods agree well with each other across most of the parameter space. Minor deviations are observed only when the planet is located close to the disk's inner edge (Panel (b2) and (d2)). Additionally, we plot the warp angles derived from $N$-body simulations. The initial simulation data points (solid circles) follow the secular theory curves, confirming the validity of the analytical approach. The results at 17~Myr (open circles) consistently exhibit the damping trend discussed in the main text.}

\bibliography{reference}{}

\begin{thebibliography}{}
\expandafter\ifx\csname natexlab\endcsname\relax\def\natexlab#1{#1}\fi
\providecommand{\url}[1]{\href{#1}{#1}}
\providecommand{\dodoi}[1]{doi:~\href{http://doi.org/#1}{\nolinkurl{#1}}}
\providecommand{\doeprint}[1]{\href{http://ascl.net/#1}{\nolinkurl{http://ascl.net/#1}}}
\providecommand{\doarXiv}[1]{\href{https://arxiv.org/abs/#1}{\nolinkurl{https://arxiv.org/abs/#1}}}

\bibitem[{{Astropy Collaboration} {et~al.}(2013){Astropy Collaboration}, Robitaille, Tollerud, Greenfield, Droettboom, Bray, Aldcroft, Davis, Ginsburg, {Price-Whelan}, Kerzendorf, Conley, Crighton, Barbary, Muna, Ferguson, Grollier, Parikh, Nair, Unther, Deil, Woillez, Conseil, Kramer, Turner, Singer, Fox, Weaver, Zabalza, Edwards, Azalee~Bostroem, Burke, Casey, Crawford, Dencheva, Ely, Jenness, Labrie, Lim, Pierfederici, Pontzen, Ptak, Refsdal, Servillat, \& Streicher}]{astropycollaboration13}
{Astropy Collaboration}, Robitaille, T.~P., Tollerud, E.~J., {et~al.} 2013, Astronomy and Astrophysics, 558, A33, \dodoi{10.1051/0004-6361/201322068}

\bibitem[{{Astropy Collaboration} {et~al.}(2018){Astropy Collaboration}, {Price-Whelan}, Sip{\H o}cz, G{\"u}nther, Lim, Crawford, Conseil, Shupe, Craig, Dencheva, Ginsburg, VanderPlas, Bradley, {P{\'e}rez-Su{\'a}rez}, {de Val-Borro}, Aldcroft, Cruz, Robitaille, Tollerud, Ardelean, Babej, Bach, Bachetti, Bakanov, Bamford, Barentsen, Barmby, Baumbach, Berry, Biscani, Boquien, Bostroem, Bouma, Brammer, Bray, Breytenbach, Buddelmeijer, Burke, Calderone, Cano~Rodr{\'i}guez, Cara, Cardoso, Cheedella, Copin, Corrales, Crichton, D'Avella, Deil, Depagne, Dietrich, Donath, Droettboom, Earl, Erben, Fabbro, Ferreira, Finethy, Fox, Garrison, Gibbons, Goldstein, Gommers, Greco, Greenfield, Groener, Grollier, Hagen, Hirst, Homeier, Horton, Hosseinzadeh, Hu, Hunkeler, Ivezi{\'c}, Jain, Jenness, Kanarek, Kendrew, Kern, Kerzendorf, Khvalko, King, Kirkby, Kulkarni, Kumar, Lee, Lenz, Littlefair, Ma, Macleod, Mastropietro, McCully, Montagnac, Morris, Mueller, Mumford, Muna, Murphy, Nelson, Nguyen, Ninan, N{\"o}the, Ogaz, Oh,
  Parejko, Parley, Pascual, Patil, Patil, Plunkett, Prochaska, Rastogi, Reddy~Janga, Sabater, Sakurikar, Seifert, Sherbert, {Sherwood-Taylor}, Shih, Sick, Silbiger, Singanamalla, Singer, Sladen, Sooley, Sornarajah, Streicher, Teuben, Thomas, Tremblay, Turner, Terr{\'o}n, {van Kerkwijk}, {de la Vega}, Watkins, Weaver, Whitmore, Woillez, Zabalza, \& {Astropy Contributors}}]{astropycollaboration18}
{Astropy Collaboration}, {Price-Whelan}, A.~M., Sip{\H o}cz, B.~M., {et~al.} 2018, The Astronomical Journal, 156, 123, \dodoi{10.3847/1538-3881/aabc4f}

\bibitem[{Augereau {et~al.}(2001)Augereau, Nelson, Lagrange, Papaloizou, \& Mouillet}]{augereau01}
Augereau, J.~C., Nelson, R.~P., Lagrange, A.~M., Papaloizou, J. C.~B., \& Mouillet, D. 2001, Astronomy \& Astrophysics, 370, 447, \dodoi{10.1051/0004-6361:20010199}

\bibitem[{Batygin(2018)}]{batygin18}
Batygin, K. 2018, Monthly Notices of the Royal Astronomical Society, 475, 5070, \dodoi{10.1093/mnras/sty162}

\bibitem[{Batygin {et~al.}(2011)Batygin, Morbidelli, \& Tsiganis}]{batygin11}
Batygin, K., Morbidelli, A., \& Tsiganis, K. 2011, Astronomy \& Astrophysics, 533, A7, \dodoi{10.1051/0004-6361/201117193}

\bibitem[{Brady {et~al.}(2023)Brady, {Faramaz-Gorka}, Bryden, \& Ertel}]{brady23}
Brady, M.~T., {Faramaz-Gorka}, V., Bryden, G., \& Ertel, S. 2023, The Astrophysical Journal, 954, 14, \dodoi{10.3847/1538-4357/ace9bb}

\bibitem[{Chen {et~al.}(2014)Chen, Mittal, Kuchner, Forrest, Lisse, Manoj, Sargent, \& Watson}]{chen14}
Chen, C.~H., Mittal, T., Kuchner, M., {et~al.} 2014, The Astrophysical Journal Supplement Series, 211, 25, \dodoi{10.1088/0067-0049/211/2/25}

\bibitem[{Crotts \& Matthews(2024)}]{crotts24}
Crotts, K.~A., \& Matthews, B.~C. 2024, The Astrophysical Journal, 975, 136, \dodoi{10.3847/1538-4357/ad7b28}

\bibitem[{{CSST Collaboration} {et~al.}(2025){CSST Collaboration}, Gong, Miao, Zhan, Li, Shangguan, Li, Liu, Chen, Yuan, Zhou, Liu, Yu, Ji, Qi, Liu, Dai, Wang, Zheng, Hao, Dou, Ao, Lin, Zhang, Wang, Sun, Li, Li, Xu, Li, Li, Wu, Zhang, Wang, Bai, Cai, Cai, Cao, Chan, Chang, Chen, Chen, Chen, Chen, Cui, Dong, Du, Duan, Fan, Fan, Fan, Fan, Fang, Fu, Fu, Fu, Gao, Gu, Gu, Guo, Han, Hu, Huang, Ho, Jiang, Jiang, Jing, Kang, Kong, Li, Li, Li, Li, Li, Li, Liao, Lin, Liu, Liu, Liu, Liu, Mao, Mao, Meng, Pang, Peng, Peng, Shan, Shen, Shen, Shen, Shi, Shi, Tan, Tian, Wang, Wang, Wang, Wang, Wu, Wu, Wu, Xu, Xue, Xue, Yang, Yang, Yao, Yuan, Yuan, Zhang, Zhang, Zhang, Zhang, Zhang, Zhao, Zhao, Zhong, Zhong, Zhou, Zhu, \& Zu}]{csst25}
{CSST Collaboration}, Gong, Y., Miao, H., {et~al.} 2025, Introduction to the Chinese Space Station Survey Telescope (CSST),  arXiv, \dodoi{10.48550/arXiv.2507.04618}

\bibitem[{Dawson {et~al.}(2011)Dawson, {Murray-Clay}, \& Fabrycky}]{dawson11}
Dawson, R.~I., {Murray-Clay}, R.~A., \& Fabrycky, D.~C. 2011, The Astrophysical Journal, 743, L17, \dodoi{10.1088/2041-8205/743/1/L17}

\bibitem[{Dou {et~al.}(2025)Dou, Zhang, Zhao, Xu, Wu, Wang, Yuan, Kong, Zhu, Niu, Lv, Qi, Jiang, Chen, Guo, Wang, Lin, Zheng, Guo, Li, Xu, Wu, Wen, Miao, Lv, \& Li}]{dou25}
Dou, J., Zhang, X., Zhao, G., {et~al.} 2025, CPI-C: Cool Planet Imaging Coronagraph on Chinese Space Station Survey Telescope,  arXiv, \dodoi{10.48550/arXiv.2512.11292}

\bibitem[{Esposito {et~al.}(2020)Esposito, Kalas, Fitzgerald, {Millar-Blanchaer}, Duch{\^e}ne, Patience, Hom, Perrin, De~Rosa, Chiang, Czekala, Macintosh, Graham, Ansdell, Arriaga, Bruzzone, Bulger, Chen, Cotten, Dong, Draper, Follette, Hung, Lopez, Matthews, Mazoyer, Metchev, Rameau, Ren, Rice, Song, Stahl, Wang, Wolff, Zuckerman, Ammons, Bailey, Barman, Chilcote, Doyon, Gerard, Goodsell, Greenbaum, Hibon, Hinkley, Ingraham, Konopacky, Maire, Marchis, Marley, Marois, Nielsen, Oppenheimer, Palmer, Poyneer, Pueyo, Rajan, Rantakyr{\"o}, Ruffio, Savransky, Schneider, Sivaramakrishnan, Soummer, Thomas, \& {Ward-Duong}}]{esposito20}
Esposito, T.~M., Kalas, P., Fitzgerald, M.~P., {et~al.} 2020, The Astronomical Journal, 160, 24, \dodoi{10.3847/1538-3881/ab9199}

\bibitem[{Faramaz {et~al.}(2019)Faramaz, Krist, Stapelfeldt, Bryden, Mamajek, Matr{\`a}, Booth, Flaherty, Hales, Hughes, Bayo, Casassus, Cuadra, Olofsson, Su, \& Wilner}]{faramaz19}
Faramaz, V., Krist, J., Stapelfeldt, K.~R., {et~al.} 2019, The Astronomical Journal, 158, 162, \dodoi{10.3847/1538-3881/ab3ec1}

\bibitem[{Grimm \& Stadel(2014)}]{grimm14}
Grimm, S.~L., \& Stadel, J.~G. 2014, The Astrophysical Journal, 796, 23, \dodoi{10.1088/0004-637X/796/1/23}

\bibitem[{Grimm \& Stadel(2018)}]{grimm18}
---. 2018, Astrophysics Source Code Library, ascl:1812.014

\bibitem[{Grimm {et~al.}(2022)Grimm, Stadel, Brasser, Meier, \& Mordasini}]{grimm22}
Grimm, S.~L., Stadel, J.~G., Brasser, R., Meier, M. M.~M., \& Mordasini, C. 2022, The Astrophysical Journal, 932, 124, \dodoi{10.3847/1538-4357/ac6dd2}

\bibitem[{Hales {et~al.}(2022)Hales, Marino, Sheehan, Ulloa, P{\'e}rez, Matr{\`a}, Kral, Wyatt, Dent, \& Carpenter}]{hales22}
Hales, A.~S., Marino, S., Sheehan, P.~D., {et~al.} 2022, The Astrophysical Journal, 940, 161, \dodoi{10.3847/1538-4357/ac9cd3}

\bibitem[{Harris {et~al.}(2020)Harris, Millman, van~der Walt, Gommers, Virtanen, Cournapeau, Wieser, Taylor, Berg, Smith, Kern, Picus, Hoyer, van Kerkwijk, Brett, Haldane, del R{\'{i}}o, Wiebe, Peterson, G{\'{e}}rard-Marchant, Sheppard, Reddy, Weckesser, Abbasi, Gohlke, \& Oliphant}]{harris20}
Harris, C.~R., Millman, K.~J., van~der Walt, S.~J., {et~al.} 2020, Nature, 585, 357, \dodoi{10.1038/s41586-020-2649-2}

\bibitem[{Hom {et~al.}(2024)Hom, Patience, Chen, Duch{\^e}ne, Mazoyer, {Millar-Blanchaer}, Esposito, Kalas, Crotts, Gonzales, Kolokolova, Lewis, Matthews, Rice, Weinberger, Wilner, Wolff, Bruzzone, Choquet, Debes, De~Rosa, Donaldson, Draper, Fitzgerald, Hines, Hinkley, Hughes, L{\'o}pez, Marchis, Metchev, {Moro-Martin}, Nesvold, Nielsen, Oppenheimer, Padgett, Perrin, Pueyo, Rantakyr{\"o}, Ren, Schneider, Soummer, Song, \& Stark}]{hom24}
Hom, J., Patience, J., Chen, C.~H., {et~al.} 2024, Monthly Notices of the Royal Astronomical Society, 528, 6959, \dodoi{10.1093/mnras/stae368}

\bibitem[{Hughes {et~al.}(2018)Hughes, Duch{\^e}ne, \& Matthews}]{hughes18}
Hughes, A.~M., Duch{\^e}ne, G., \& Matthews, B.~C. 2018, Annual Review of Astronomy and Astrophysics, 56, 541, \dodoi{10.1146/annurev-astro-081817-052035}

\bibitem[{Hunter(2007)}]{hunter07}
Hunter, J.~D. 2007, Computing in Science \& Engineering, 9, 90, \dodoi{10.1109/MCSE.2007.55}

\bibitem[{Kalas {et~al.}(2005)Kalas, Graham, \& Clampin}]{kalas05}
Kalas, P., Graham, J.~R., \& Clampin, M. 2005, Nature, 435, 1067, \dodoi{10.1038/nature03601}

\bibitem[{Kasdin {et~al.}(2020)Kasdin, Bailey, Mennesson, Zellem, Ygouf, Rhodes, Luchik, Zhao, Riggs, Seo, Krist, Kern, Tang, Nemati, Groff, Zimmerman, Macintosh, Turnbull, Debes, Douglas, \& Lupu}]{kasdin20}
Kasdin, N.~J., Bailey, V.~P., Mennesson, B., {et~al.} 2020, in Space Telescopes and Instrumentation 2020: Optical, Infrared, and Millimeter Wave, 194, \dodoi{10.1117/12.2562997}

\bibitem[{Kasper {et~al.}(2015)Kasper, Apai, Wagner, \& Robberto}]{kasper15}
Kasper, M., Apai, D., Wagner, K., \& Robberto, M. 2015, The Astrophysical Journal, 812, L33, \dodoi{10.1088/2041-8205/812/2/L33}

\bibitem[{Krivov \& Wyatt(2021)}]{krivov21}
Krivov, A.~V., \& Wyatt, M.~C. 2021, Monthly Notices of the Royal Astronomical Society, 500, 718, \dodoi{10.1093/mnras/staa2385}

\bibitem[{Lagrange {et~al.}(2025)Lagrange, Wilkinson, M{\^a}lin, Boccaletti, Perrot, Matr{\`a}, Combes, Beust, Rouan, Chomez, Milli, Charnay, Mazevet, Flasseur, Olofsson, Bayo, Kral, Carter, Crotts, Delorme, Chauvin, Thebault, Rubini, Kiefer, Radcliffe, Mazoyer, Bodrito, Stasevic, \& Langlois}]{lagrange25a}
Lagrange, A.-M., Wilkinson, C., M{\^a}lin, M., {et~al.} 2025, Nature, 642, 905, \dodoi{10.1038/s41586-025-09150-4}

\bibitem[{Melton \& Batygin(2021)}]{melton21}
Melton, W., \& Batygin, K. 2021, Monthly Notices of the Royal Astronomical Society, 502, 3955, \dodoi{10.1093/mnras/stab344}

\bibitem[{Mouillet {et~al.}(1997)Mouillet, Larwood, Papaloizou, \& Lagrange}]{mouillet97}
Mouillet, D., Larwood, J.~D., Papaloizou, J. C.~B., \& Lagrange, A.~M. 1997, Monthly Notices of the Royal Astronomical Society, 292, 896, \dodoi{10.1093/mnras/292.4.896}

\bibitem[{Murray \& Dermott(1999)}]{murray99}
Murray, C.~D., \& Dermott, S.~F. 1999, Solar System Dynamics (Cambridge ; New York: Cambridge University Press)

\bibitem[{Nederlander {et~al.}(2021)Nederlander, Hughes, Fehr, Flaherty, Su, Mo{\'o}r, Chiang, Andrews, Wilner, \& Marino}]{nederlander21}
Nederlander, A., Hughes, A.~M., Fehr, A.~J., {et~al.} 2021, The Astrophysical Journal, 917, 5, \dodoi{10.3847/1538-4357/abdd32}

\bibitem[{Nesvold \& Kuchner(2015)}]{nesvold15}
Nesvold, E.~R., \& Kuchner, M.~J. 2015, The Astrophysical Journal, 815, 61, \dodoi{10.1088/0004-637X/815/1/61}

\bibitem[{Poblete {et~al.}(2023)Poblete, L{\"o}hne, Pearce, \& Sefilian}]{poblete23}
Poblete, P.~P., L{\"o}hne, T., Pearce, T.~D., \& Sefilian, A.~A. 2023, Monthly Notices of the Royal Astronomical Society, 526, 2017, \dodoi{10.1093/mnras/stad2827}

\bibitem[{Sefilian(2024)}]{sefilian24}
Sefilian, A.~A. 2024, The Astrophysical Journal, 966, 140, \dodoi{10.3847/1538-4357/ad32d1}

\bibitem[{Sefilian {et~al.}(2025)Sefilian, Kratter, Wyatt, Petrovich, Thébault, Malhotra, \& Faramaz-Gorka}]{sefilian25}
Sefilian, A.~A., Kratter, K.~M., Wyatt, M.~C., {et~al.} 2025, staf1555, \dodoi{10.1093/mnras/staf1555}

\bibitem[{Sefilian {et~al.}(2021)Sefilian, Rafikov, \& Wyatt}]{sefilian21}
Sefilian, A.~A., Rafikov, R.~R., \& Wyatt, M.~C. 2021, The Astrophysical Journal, 910, 13, \dodoi{10.3847/1538-4357/abda46}

\bibitem[{Sefilian {et~al.}(2023)Sefilian, Rafikov, \& Wyatt}]{sefilian23}
---. 2023, The Astrophysical Journal, 954, 100, \dodoi{10.3847/1538-4357/ace68e}

\bibitem[{Stasevic {et~al.}(2023)Stasevic, Milli, Mazoyer, Lagrange, Bonnefoy, {Faramaz-Gorka}, M{\'e}nard, Boccaletti, Choquet, Shuai, Olofsson, Chomez, Ren, Rubini, Desgrange, Gratton, Chauvin, Vigan, \& Matthews}]{stasevic23}
Stasevic, S., Milli, J., Mazoyer, J., {et~al.} 2023, Astronomy \& Astrophysics, 678, A8, \dodoi{10.1051/0004-6361/202346720}

\bibitem[{{Ulubay-Siddiki} {et~al.}(2009){Ulubay-Siddiki}, Gerhard, \& Arnaboldi}]{ulubay-siddiki09}
{Ulubay-Siddiki}, A., Gerhard, O., \& Arnaboldi, M. 2009, Monthly Notices of the Royal Astronomical Society, 398, 535, \dodoi{10.1111/j.1365-2966.2009.15089.x}

\bibitem[{Virtanen {et~al.}(2020)Virtanen, Gommers, Oliphant, Haberland, Reddy, Cournapeau, Burovski, Peterson, Weckesser, Bright, {van der Walt}, Brett, Wilson, Millman, Mayorov, Nelson, Jones, Kern, Larson, Carey, Polat, Feng, Moore, {VanderPlas}, Laxalde, Perktold, Cimrman, Henriksen, Quintero, Harris, Archibald, Ribeiro, Pedregosa, {van Mulbregt}, \& {SciPy 1.0 Contributors}}]{scipy2020}
Virtanen, P., Gommers, R., Oliphant, T.~E., {et~al.} 2020, Nature Methods, 17, 261, \dodoi{10.1038/s41592-019-0686-2}

\bibitem[{Wyatt(2006)}]{wyatt06}
Wyatt, M.~C. 2006, The Astrophysical Journal, 639, 1153, \dodoi{10.1086/499487}

\bibitem[{Wyatt {et~al.}(1999)Wyatt, Dermott, Telesco, Fisher, Grogan, Holmes, \& Pina}]{wyatt99}
Wyatt, M.~C., Dermott, S.~F., Telesco, C.~M., {et~al.} 1999, The Astrophysical Journal, 527, 918, \dodoi{10.1086/308093}

\bibitem[{Zanazzi \& Lai(2018)}]{zanazzi18}
Zanazzi, J.~J., \& Lai, D. 2018, Monthly Notices of the Royal Astronomical Society, 473, 603, \dodoi{10.1093/mnras/stx2375}

\end{thebibliography}
\bibliographystyle{aasjournal}



\end{document}